\documentclass[runningheads]{llncs}

\usepackage{cite}
\usepackage{url}
\usepackage{amsmath,amsfonts} 
\usepackage{float}
\usepackage{graphicx}
\usepackage{algorithm}
\usepackage{algpseudocode}
\usepackage{siunitx}
\usepackage[labelfont=bf]{caption}
\usepackage{subcaption}
\usepackage{tikz}
\usepackage{pgfplots}
\usepackage{epstopdf}
\usepackage{multirow}
\usepackage[font=scriptsize]{caption}
\usepackage{soul}
\usepackage{color, colortbl}
\usepackage{array}
\usepackage[T1]{fontenc}
\usepackage[utf8]{inputenc}
\usepackage{mathptmx}
\usepackage{listings}

\usepackage{enumitem}
\usepackage{hyperref}

\begin{document}

\title{A Secure Contained Testbed For Analyzing IoT Botnets} 
\author{Ayush Kumar \and Teng Joon Lim}
\authorrunning{A. Kumar et al.} 
\institute{Department of Electrical and Computer Engineering, \\ 
National University of Singapore, Singapore 119077 \\
			\email{ayush.kumar@u.nus.edu}, \email{eleltj@nus.edu.sg}}
\maketitle

\begin{abstract}
Many security issues have come to the fore with the increasingly widespread adoption of Internet-of-Things (IoT) devices. The Mirai attack on Dyn DNS service, in which vulnerable IoT devices such as IP cameras, DVRs and routers were infected and used to  propagate large-scale DDoS attacks, is one of the more prominent recent examples. IoT botnets, consisting of hundreds-of-thousands of bots, are currently present ``in-the-wild''  at least and are only expected to grow in the future, with the potential to cause significant network downtimes and financial losses to network companies. We propose, therefore, to build testbeds for evaluating IoT botnets and design suitable mitigation techniques against them. A DETERlab-based IoT botnet testbed is presented in this work. The testbed is built in a secure contained environment and includes ancillary services such as DHCP, DNS as well as botnet infrastructure including CnC and scanListen/loading servers. Developing an IoT botnet testbed presented us with some unique challenges which are different from those encountered in non-IoT botnet testbeds and we highlight them in this paper. Further, we point out the important features of our testbed and illustrate some of its capabilities through experimental results.
\end{abstract}

\section{Introduction}
The Internet of things (IoT)\cite{iotsurvey} refers to the network of low-power, limited processing capability sensing devices which can  send/receive data to/from other devices using wireless technologies such as RFID (Radio Frequency Identification), Zigbee, WiFi, Bluetooth, cellular etc. IoT devices are being deployed in a number of applications such as wearables, home automation, smart grids, environmental monitoring, infrastructure management, industrial automation, agricultural automation, healthcare and smart cities. Some of the popular platforms for IoT are Samsung SmartThings (consumer IoT for device management) and Amazon Web Services IoT, Microsoft Azure IoT, Google Cloud Platform (enterprise IoT for cloud storage and data analytics). The number of IoT devices deployed globally by 2020 is expected to be in the range of 20-30 billion \cite{iotforecast}. The number of devices has been increasing steadily (albeit at a slower rate than some earlier generous predictions), and this trend is expected to hold in the future.

IoT devices are being increasingly targeted by hackers using malware (malicious software) as they are easier to infect than conventional computers for the following reasons\cite{iotsecsurvey1,iotsecsurvey2,iotsecsurvey3}:
\begin{itemize}
\item There are many legacy IoT devices connected to the Internet with no security updates.
\item Security is given a low priority within the development cycle of IoT devices.
\item Implementing conventional cryptography in IoT devices is computationally expensive due to processing power and memory constraints.
\item Many IoT devices have weak login credentials either provided by the manufacturer or configured by users. 
\item IoT device manufacturers sometimes leave \textit{backdoors} (such as an open port) to provide support for the device remotely.
\item Often, consumer IoT devices are connected to the Internet without going through a firewall.  
\end{itemize}  
In a widely publicized attack, the IoT malware \textit{Mirai} was used to propagate the biggest DDoS (Distributed Denial-of-Service) attack on record on October 21, 2016. The attack targeted the Dyn DNS (Domain Name Service) servers \cite{miraiattack} and generated an attack throughput of the order of 1.2 Tbps. It disabled major internet services such as Amazon, Twitter and Netflix. The attackers had infected IoT devices such as IP cameras and DVR recorders with Mirai, thereby creating an army of bots (botnet) to take part in the DDoS attack. Apart from Mirai, there are other IoT malware which operate using a similar brute force technique of scanning random IP addresses for open ports and attempting to login using a built-in dictionary of commonly used credentials. BASHLITE \cite{bashlite}, Remaiten \cite{remaiten}, and Hajime \cite{hajime} are some examples of these IoT malware. 
 
Bots compromised by Mirai or similar IoT malware can be used for DDoS attacks, phishing and spamming \cite{phishspam}. These attacks can cause network downtime for long periods which may lead to financial loss to network companies, and leak users' confidential data. McAfee reported in April 2017\cite{mcafee} that about 2.5 million IoT devices were infected by Mirai in late 2016. Bitdefender mentioned in its blog in September 2017\cite{bitdef} that researchers had estimated at least 100,000 devices infected by Mirai or similar malware revealed daily through telnet scanning telemetry data.  Further, many of the infected devices are expected to remain infected for a long time. Therefore, there is a substantial motivation for studying these IoT botnets so as to characterize, detect and develop effective countermeasures against them. 

A comprehensive IoT botnet testbed environment will be quite useful for researchers working towards this goal. Further, the testbed can be used to generate useful ground truth data for researchers to evaluate the effectiveness of their proposed IoT botnet detection algorithms. As pointed out in \cite{3000bot}, botnet emulation studies in controlled laboratory environments have a number of advantages (e.g. closeness to real-world botnets, greater degree of safety and control over experiment environment, fewer legal and ethical issues) when it comes to botnet research compared to analytical modelling, simulation studies, botnet binary reverse engineering and in-the-wild botnet analysis.

In this work, we present a testbed environment built using the National Cybersecurity R\&D Lab (NCL)\cite{ncl} infrastructure and consisting of physical and virtual machines (VMs) which can be used to study Mirai-like IoT botnets. As pointed out later, the same testbed can be modified to work with other IoT botnets as well. The testbed includes ancillary services such as Domain Name Service (DNS), Dynamic Host Configuration Protocol (DHCP), botnet infrastructure which includes Command-and-Control (CnC) and scanListen/loading servers, and a secure contained environment for testing. The last two services are provided by DETERLab \cite{deterlab}, which provides the software stack on which the NCL testbed is based. A secure contained environment is especially critical since many of the botnets are capable of further infecting devices connected to the Internet. This testbed is mainly targeted at cyber security researchers who can easily and quickly bootstrap IoT botnet experiments on our testbed. To the best of our knowledge, this is the first testbed to emulate the full behavior of an IoT botnet. In the subsequent sections, we review published research on botnet testbeds. We also elaborate on our experience setting up the various components of the testbed and the related challenges that we faced. Finally, we include some experimental results from our testbed that illustrate its capabilities. 

\section{Related Work}
There have been several research works on the study of botnets through testbeds. Barford et al\cite{mesocosms} have presented a toolkit for Emulab-enabled testbeds (such as DETERlab) called Botnet Evaluation Environment (BEE), designed to provide bots and botnets for experimentation in a scalabale and secure/self-contained environment. BEE includes a library of OS/bot images to be run on physical and virtual machines as well as a set of services and tools required for botnet evaluation such as DHCP, DNS, IRC, VM monitors and honeypots. \cite{slingbot} has proposed a composable botnet framework called SLINGbot which can be used to construct botnets with different (Command and Control) C2 structures (including potential future botnet C2 structures), simulate botnet traffic, characterize it and develop botnet defense techniques. 

In \cite{3000bot}, the authors have argued for isolated in-the-lab at-scale botnet experimentation and evaluated an emulated 3000-node Waledac botnet. Further, they have also validated a defense technique (using sybil attack) against the botnet. \cite{largescale} has argued for large-scale botnet emulation that is at par with actual botnets (hundreds of thousands to millions of bots) and potentially discover issues that show up at that scale. The authors have presented a prototype testbed with 600,000 Linux VMs or 62,000 Windows 7 VMs on a 520 node computing cluster through over-budgeting of processor and memory resources and other techniques. Since few  works have paid attention to containment in botnet emulation, a malware execution farm, \textit{GQ} has been presented in \cite{gqcontain} which focuses on methodical development of explicit containment policies for malware measurement and analysis. 

ElSheikh et al.\cite{botgen} have addressed the lack of publicly available botnet research datasets by creating an in-lab botnet experimentation testbed in a contained environment and using it to generate botnet datasets. \cite{httpbotnet} has implemented an HTTP (Hypertext Transfer Protocol)-based botnet testbed for performing HTTP GET flooding attacks against web servers. The authors have also provided real-time data sets (including http bot traces) for researchers to evaluate such botnets and proposed countermeasures against DDoS attacks from those botnets. Finally, in \cite{vnetwork}, the authors have presented a controlled network environment based on VMware virtualization technology, called V-Network, for analyzing network worm propagation patterns and evaluating countermeasure systems. 

This paper makes an important contribution since there has been no work till now on creating IoT botnet testbeds emulating the full behavior of IoT malware. Our testbed is similar to \cite{mesocosms} in the sense that it is also based on DETERlab and provides services such as DNS, DHCP, CnC etc. but the challenges encountered by us (as pointed out in section \ref{testbedchallenges}) were different because the testbed in \cite{mesocosms} was focused on PC-based bots and not IoT bots. In comparison to \cite{3000bot} and \cite{largescale}, our current focus is not on scalability since we can increase the number of bots in our testbed easily by installing more VMs per physical machine and adding more physical machines. However, we intend to look into their techniques in the future to increase the number of bots in our testbed to more accurately represent real-world IoT botnets. We are in the process of collecting datasets for the IoT botnet research community, similar to \cite{botgen}, and hope to release them in near future. Out testbed is not restricted to specific attacks such as HTTP flooding\cite{httpbotnet} and we also use virtualization to create multiple bots on a physical machine, albeit not through VMWare hypervisor as done in \cite{vnetwork}.
\section{IoT Botnet Testbed}
In this section, we begin by giving an overview of the operation of Mirai, followed by a detailed discussion on the process of setting up various components of our IoT botnet testbed. We move on to highlight the challenges during testbed setup and how they were overcome. Finally, the major features of our testbed are pointed out.

\subsection{Overview of Mirai}
The Mirai \cite{mirai} setup consists of three major components: \textit{bot}, \textit{scanListen}/\textit{loading} server, and the \textit{CnC} (Command-and-Control) server. The \textit{CnC} server also functions as a MySQL\cite{mysql} database server. User accounts can be created in this database for customers who wish to hire DDoS-as-a-service. The database on CnC server consists of three tables: history, user and whitelist. They are assigned bots and can use them to launch a DDoS attack against their target services. There are a number of attack options available in Mirai: UDP flood, SYN flood, ACK flood, TCP stomp flood, UDP plain flood, Valve source engine specific flood, DNS resolver flood, GRE IP flood, GRE Ethernet flood and HTTP flood.

\begin{figure}[h]
\centering
\includegraphics[scale=0.4]{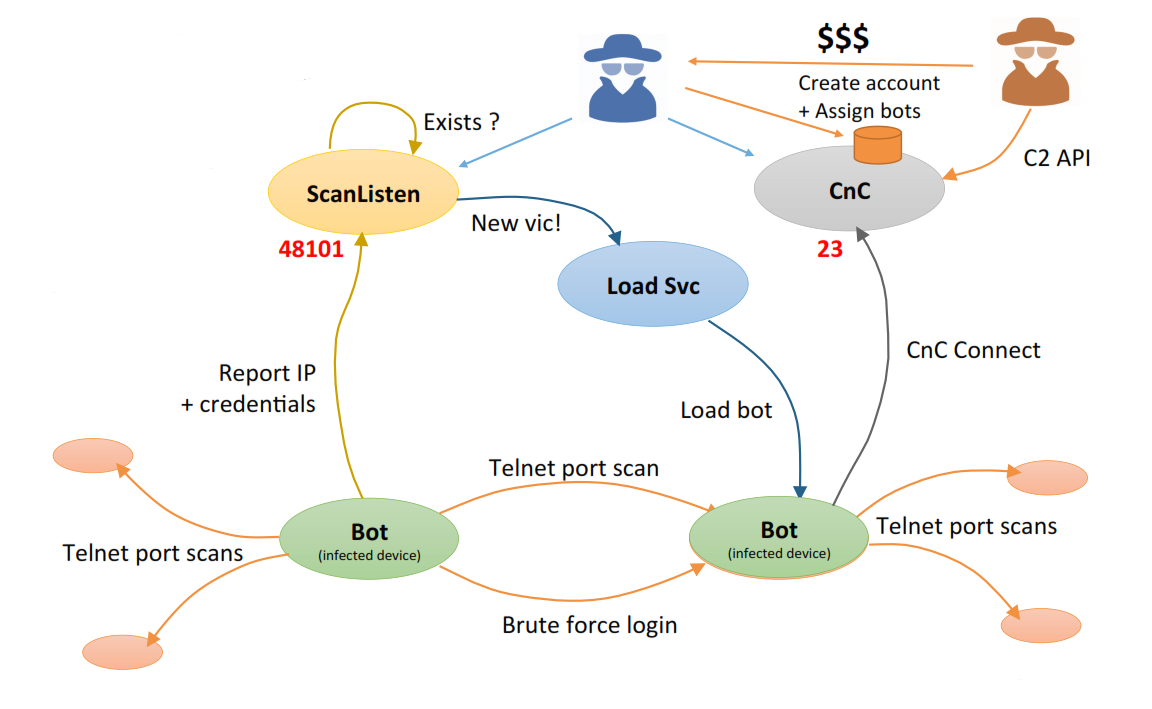}
\caption{Operation of various components of Mirai \textit{(Source: Radware \cite{miraipres})}}
\label{mirai_op}
\end{figure}

The operation of Mirai is illustrated in Fig. \ref{mirai_op}. Once an IoT device is infected with Mirai (and becomes a bot), it first attempts to connect to the listening \textit{CnC} server by resolving its domain name and opening a socket connection. Thereafter, it starts scanning the network by sending SYN packets to random IP addresses and waiting for them to respond. This process may take a long time since the bot has to go through a large number of IP addresses. Once it finds a vulnerable device with a TELNET port open, it attempts to open a socket connection to that device and emulates the TELNET protocol. Then it attempts to login using a list of default credentials and if successful, it reports the IP address of the discovered device and the working TELNET login credentials to the listening \textit{scanListen} server.
The \textit{scanListen} server sends that information to the \textit{loader} which again logs in to the discovered device using the details received from the \textit{scanListen} server. Once logged in, the \textit{loader} downloads the Mirai bot binary to that device and the new bot connects to the \textit{CnC} server and starts scanning the network.

\subsection{Installing Testbed Components}
\begin{figure}[h]
\centering
\includegraphics[scale=0.4]{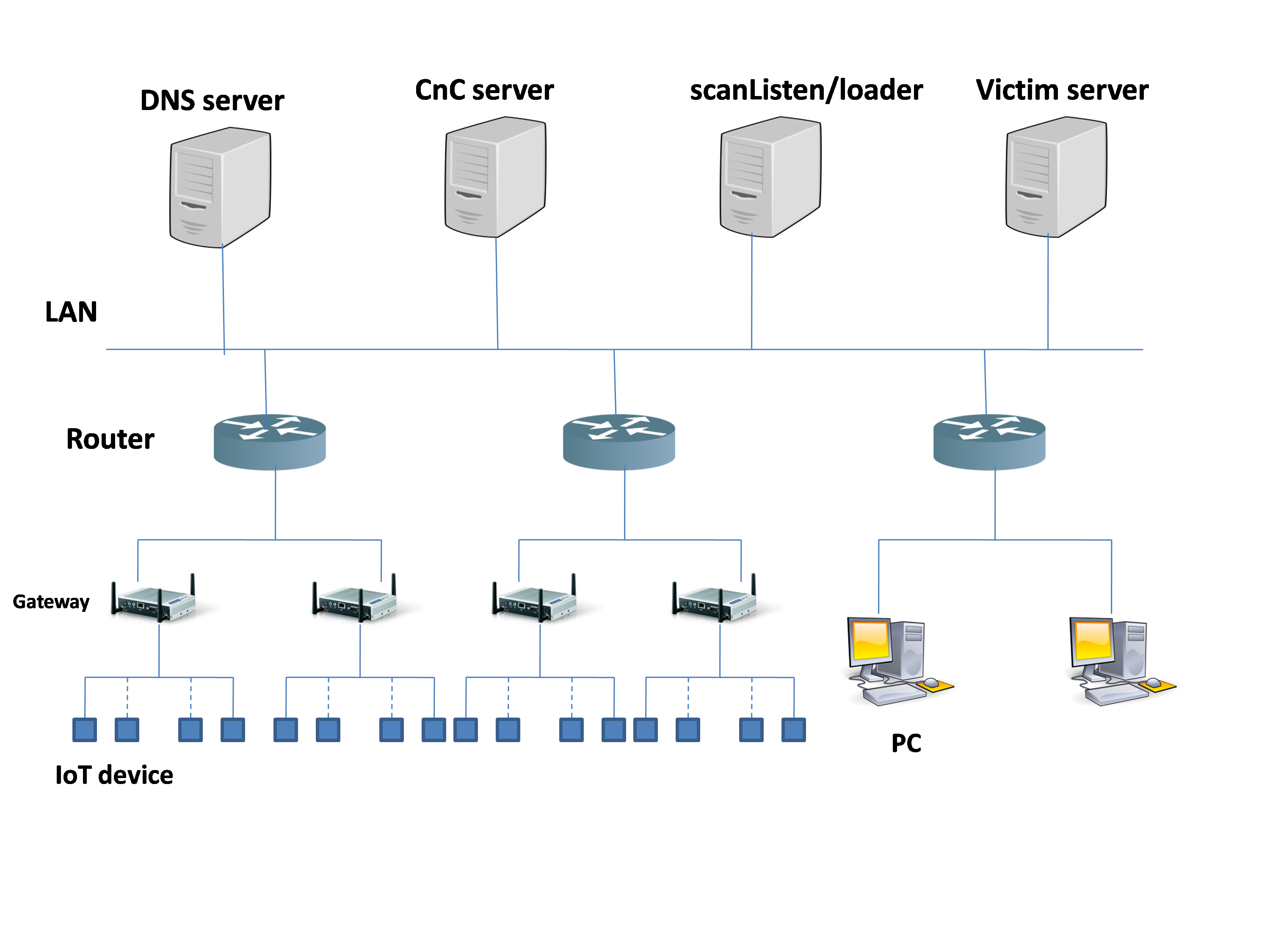}
\caption{Testbed used to simulate Mirai behavior}
\label{mirai_testbed}
\end{figure}
The testbed shown in Fig. \ref{mirai_testbed} was configured on a computing cluster isolated from the Internet. Each cluster node has two Intel Xeon E5-2620 processors, 64 GB DDR4 ECC memory and runs Ubuntu 14.04 LTS standard image. The first step was to configure a local authoritative DNS server on one of the testbed nodes because infected Mirai bots connect to CnC and scanListen servers using domain names instead of fixed IP addresses. The reason behind this design can be that the malware authors did not want the CnC/scanListen servers to be taken down if their IP addresses were identified by security personnel. If domain names are used for connection instead, the IP addresses mapped to those domain names can be easily changed if they have been identified and blocked. Originally, the Mirai source code (\textit{bot/resolv.c}) uses the Google DNS servers located at IP address \textit{8.8.8.8} for domain name-to-IP address conversion. However, since we were working on a secure isolated testbed with no outside Internet connectivity , we could not use Google DNS servers. Therefore, it became essential for us to configure a local DNS server on our testbed. We used BIND9 (Berkeley Internet Naming Daemon) to configure a local name server as a primary master, creating forward and reverse zone files. 
The next step entailed configuring a CnC server (by creating a MYSQL database) on a testbed node and the scanListen/loader on another node. 

The IoT devices used in our testbed are Raspberry Pi (RPi) devices emulated using QEMU\cite{qemu}, an open source machine emulator and hardware virtualizer. Each RPi device runs Raspbian (Wheezy version) and is configured with ARM 1176 (v6) CPU and 256MB of RAM. The CnC and scanListen server domain names and port numbers were updated in \textit{bot/table.c} file in the RPi. Since the emulated RPi devices need to connect to the CnC and scanListen servers using their domain names, the details of local DNS server were added to the \textit{/etc/resolv.conf} files in emulated devices which query the local DNS server to get the IP addresses corresponding to the domain names and then connect to those IP addresses. The codes for bot, CnC server, scanListen server and loader were compiled using the script, \textit{mirai/build.sh}. 
Since we wanted to compile the bot binaries for RPi (which runs on ARM platform) on a testbed node running on x86 platform, we also had to setup cross-compilers using the \textit{scripts/cross-compile.sh} script. 

\subsection{Challenges Encountered in Testbed Setup}
\label{testbedchallenges}
One of the major challenges was the TELNET protocol emulation in Mirai source code. When the original bot binary compiled from source code was run on an emulated RPi device, it was unable to receive the TELNET username prompt. Without the username prompt, it was impossible to try the usernames from the list of default credentials and find a working one. Hence, the first bot could never successfully infect another emulated device, thus preventing us from simulating the real propagation behavior of Mirai on our testbed. It became imperative for us to modify the code handling the TELNET protocol emulation in the original Mirai source code so that it worked on our testbed.

\begin{figure}[h]
\centering
\includegraphics[scale=0.4]{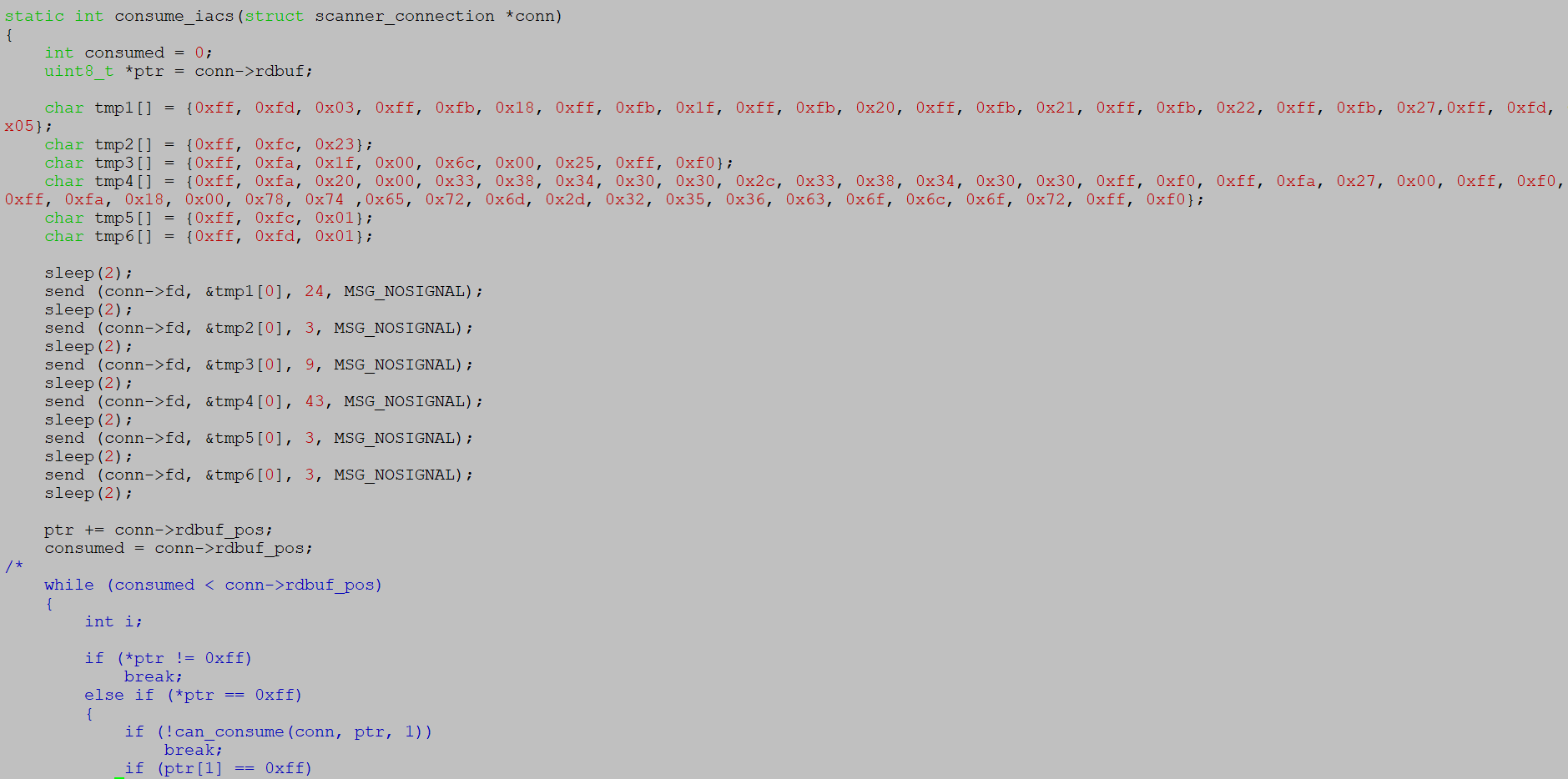}
\caption{Changes made in \textit{consume\_iacs()} function in Mirai source code}
\label{mirai-consume-iacs}
\end{figure}

\begin{figure}[h]
\centering
\includegraphics[scale=0.4]{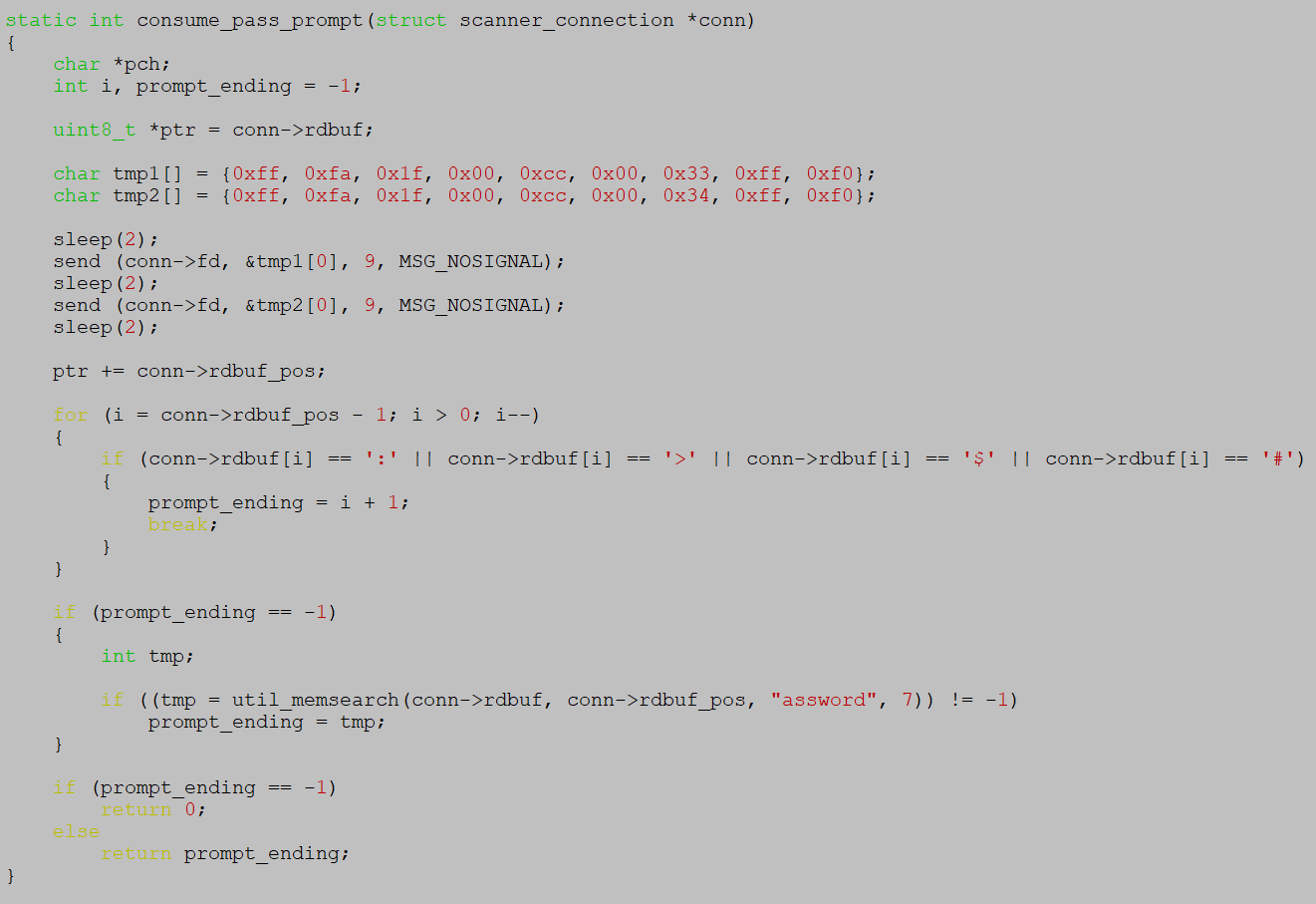}
\caption{Changes made in \textit{consume\_pass\_prompt()} function in Mirai source code}
\label{mirai-consume-pass}
\end{figure}

\begin{figure}[h]
\centering
\includegraphics[scale=0.8]{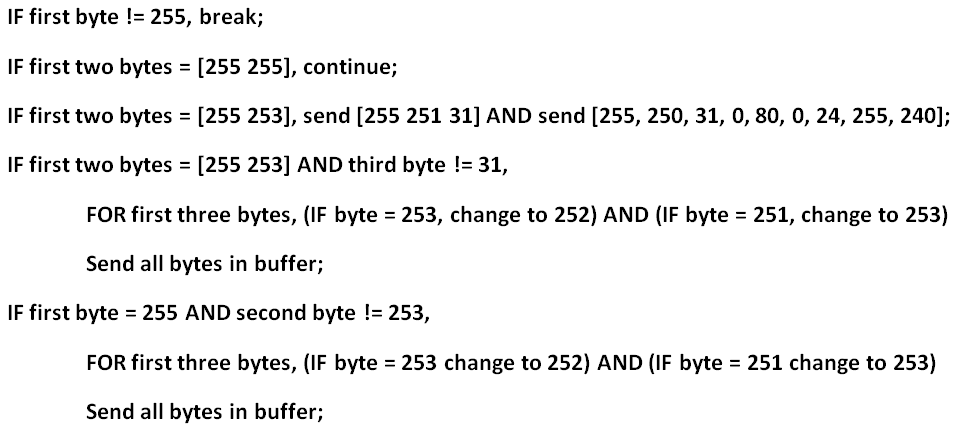}
\caption{TELNET client-server negotiation logic used in Mirai source code}
\label{mirai-telnet-neg-logic}
\end{figure}

When a TELNET client makes a connection request to a server, there is a negotiation phase where an exchange of a specific sequence of bytes takes place between the client and the server before the client is served with the username prompt. The original TELNET byte exchange logic in Mirai source code (function: \textit{consume\_iacs()}, file: \textit{bot/scanner.c}) is shown in Fig. \ref{mirai-telnet-neg-logic}. Since this logic was not working on our testbed, we created a hack and sent a TELNET connection request from one of our testbed nodes to another. The underlying byte exchange was analyzed using \textit{tcpdump} which is a very useful packet analyzer that is usually installed by default on Linux machines. We modified the code in \textit{consume\_iacs()} function to replicate the above byte exchange sequence which is depicted in Fig. \ref{mirai-consume-iacs}. Again, after the username prompt is served to the client and the username is entered, there is a byte exchange between the TELNET client and server before the password prompt is served. Therefore, we modified the code in \textit{consume\_pass\_prompt()} with the required byte sequence so that the testbed bot is served the password prompt, as shown in Fig. \ref{mirai-consume-pass}.

For the emulated RPi (on which Mirai binary is executed) to scan nearby devices, it needs to connect to the host network. By default, QEMU assigns a static IP address (10.0.2.15) to the emulated RPi if the user has not specified any network interfaces in the script that brings up the RPi, thereby leaving no connectivity between RPi and host. One of the ways to achieve this connectivity is by using bridge-TAP configuration. A network bridge refers to a device that connects two separate network segments using the same communication protocol, allowing them to operate as a single aggregate network. TAP interface is a virtual network interface that sends and receives layer 2 packets such as Ethernet frames not through a physical ``wire'', but instead by writing and reading data respectively from user-space programs. We used \textit{brctl} to configure Ethernet bridge on the host, \textit{tunctl} to create and manage TAP interfaces and \textit{libvrt} virtualization management library to automatically setup a DHCP server and bridge (\textit{virbr0}) on the host. The \textit{virbr0} comes up with a default IP address(192.168.122.1), so the IP address had to be changed for each host using \textit{virsh} utility. The TAP interface was bound to the bridge and the QEMU RPi was brought up with a network interface attached to the TAP device. Any packets sent from the RPi are now written to the TAP interface and forwarded by the attached bridge to the host network. Similarly, packets sent from the host network are forwarded through the bridge to the attached TAP interface from where they are read by RPi operating system. Since we were emulating multiple QEMU VMs on a physical node, a base Raspbian image was created in \textit{qcow2} format and using it as a backing file, we generated disk images which store only the differences from base image to avoid using multiple raw image files and save on storage space.

\subsection{Testbed Features}
The main features of our IoT botnet testbed are listed below:
\begin{itemize}
\item Using Network Simulator (NS) scripts, users can conveniently start, stop, modify and restart malware experiments instead of employing manual configuration.
\item The emulated Raspberry Pi devices are quite close to the real-world IoT devices in terms of the platform, operating system, applications and networking capabilities.
\item Our testbed can be scaled to add more IoT devices (by increasing the number of QEMU VMs per physical machine), gateways, routers and other network devices as per user simulation requirements. It also supports the integration of both physical and virtual devices.
\item The testbed topology, i.e. the way the routers, gateways and IoT devices are connected to each other, can be changed to simulate different types of networks.
\item More advanced versions or derivatives of Mirai malware can be tested on our testbed since the ancillary infrastructure required (consisting of DNS server, CnC server, scanListen/loader) is already configured. The only remaining requirement is to run the specific malware binary on the emulated IoT devices.
\item IoT malware exploiting software vulnerabilities (such as Reaper, Satori etc.) can also be tested on our testbed by running the corresponding bot binaries/compiled source codes/exploit codes on the emulated RPi devices and making a few other modifications in the ancillary services and botnet infrastructure.
\end{itemize}

\section{Experimental Results}
As shown in Fig. \ref{mirai_testbed}, the local DNS server, CnC (Command-and-Control) server, scanListen/loading server and the victim server (which is to be attacked through DDoS) are all connected to a single LAN. In practice, all the above servers are parts of different sub-networks connected to the Internet. The NCL testbed has a limitation of maximum two network interfaces per physical node, thereby restricting us from simulating more realistic networks with multiple connections. However, this restriction does not impair our testbed functions such as scanning of vulnerable devices, bot-CnC communications, infection of emulated devices and attacks launched from infected devices. These functions are are not dependent on the relative placement of DNS server, CnC server, scanListen/loading server and victim server in the network. Further connected to the same LAN are three routers, two of which are connected to IoT gateways (physical devices acting as a bridge between IoT devices and the cloud) and the third one is connected to non-IoT devices (e.g. PC). Finally, each gateway is connected to $10$ IoT devices. We chose this gateway-IoT device topology since it is used in a number of IoT deployments (such as IP cameras, smart lighting devices, wearables etc.). The IoT devices can run a video streaming server to simulate the operation of an IP camera (IoT device used in Mirai attack on Dyn). A few IoT devices were configured to have their TELNET port number $23$ open and listening for connections.

\begin{figure}[h]
\centering
\includegraphics[scale=0.8]{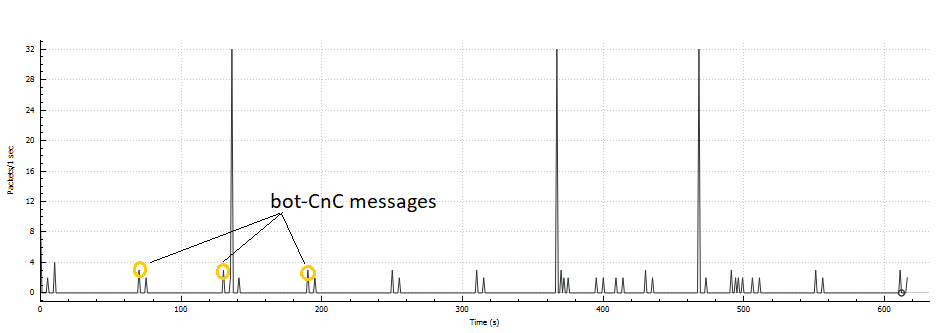}
\caption{Volume of bot-CnC traffic with time}
\label{mirai-bot-cnc-traffic}
\end{figure}

\begin{figure}[h]
\centering
\includegraphics[scale=0.6]{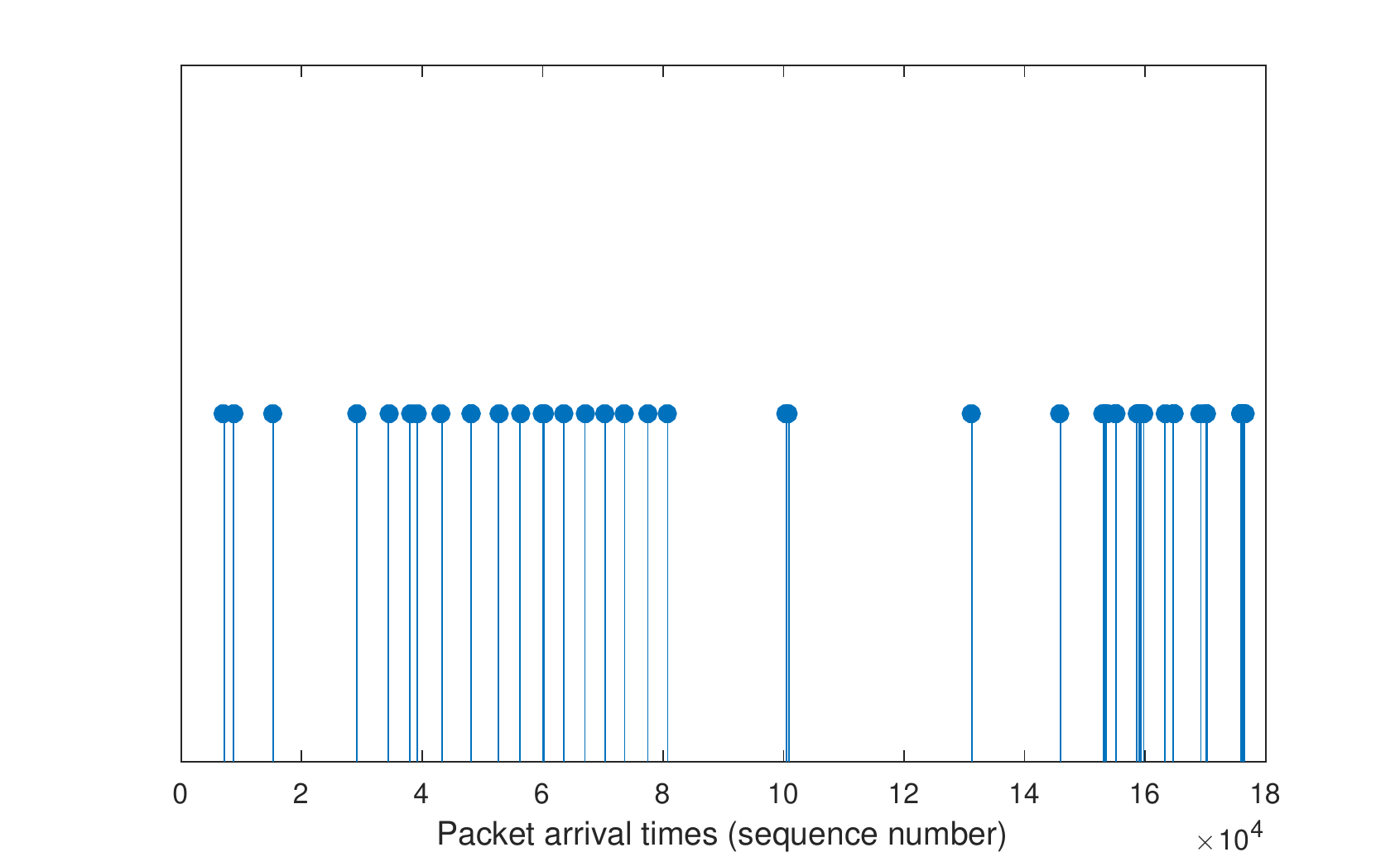}
\caption{Transmission times of bot scanning packets}
\label{scanpkt_arrivals2}
\end{figure}

We have plotted the volume of bot-CnC traffic with time for a single bot (traffic volume in packets per second) in Fig. \ref{mirai-bot-cnc-traffic}. As expected, there are busy and non-busy periods in the bot-CnC traffic. The busy period is when there is exchange of keep-alive messages (PSH+ACK) between the bot and the CnC server. Here, PSH refers to a push message and ACK refers to acknowledgement. Since the keep-alive messages are sent at regular intervals, one can observe periodic spikes in the bot-CnC traffic rate. The transmission of bot scanning packets is illustrated in Fig.\ref{scanpkt_arrivals2}. As can be observed, the scanning packets are sometimes transmitted within short intervals and at other times they are transmitted far apart, resembling a Poisson process. 

\begin{figure}[h]
\centering
\includegraphics[scale=0.6]{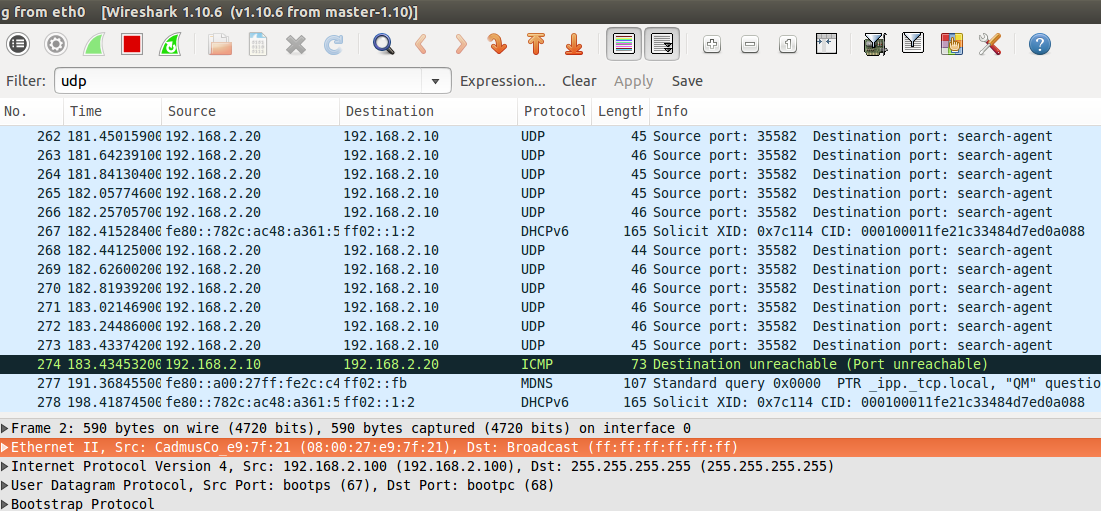}
\caption{Client UDP connection packet capture after UDP flood attack}
\label{udp-attack}
\end{figure}

In the next experiment, we commanded few infected bots to attack the victim server with UDP flood traffic which reaches a rate of 48K pps (packets per second). The attack does not wait for ICMP (Internet Control Message Protocol) reply packets (such as destination unreachable) to be sent back by the victim server. One of the non-IoT nodes in the testbed was configured as a UDP client. Once the victim server was overwhelmed, it started dropping legitimate UDP requests from the client. This can be noted from Fig. \ref{udp-attack} which depicts the client UDP connection packet capture after the attack.

\begin{figure}[h]
\centering
\includegraphics[scale=0.6]{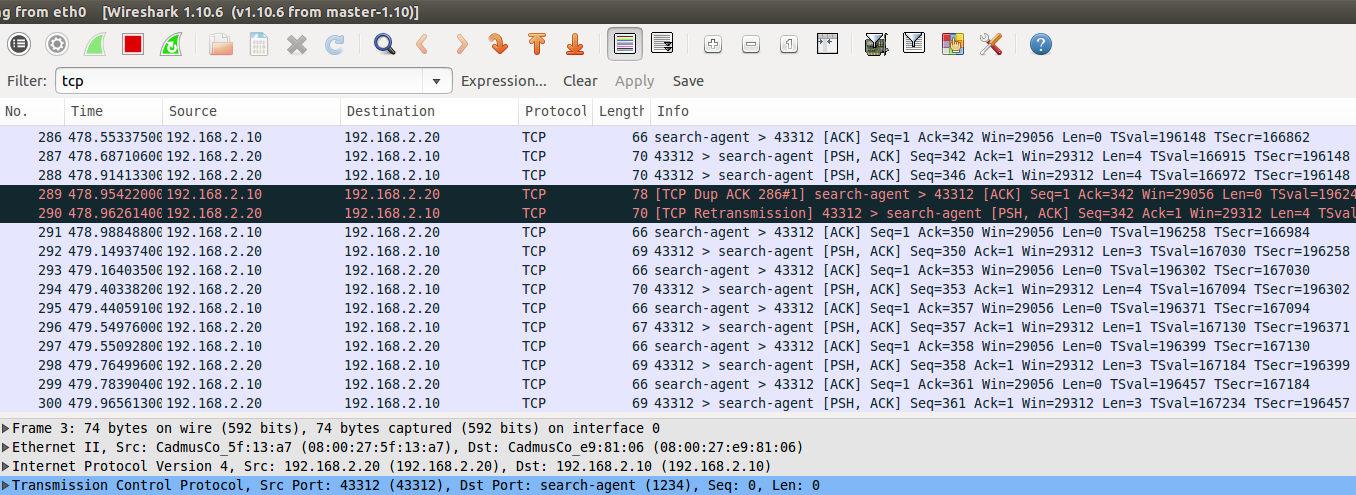}
\caption{Client TCP connection packet capture after SYN flood attack}
\label{tcp-attack}
\end{figure}

We also conducted a TCP SYN flooding attack in which $\approx$ 46K pps were transmitted by the bots without listening for SYN-ACK packets sent back by the victim server. During the attack, the victim server encountered congestion as can be inferred from Fig. \ref{tcp-attack}. It started sending out duplicate ACKs and retransmits packets multiple times which stop once the attack is over. This resulted in a denial of service to the benign TCP traffic from the client. 

\section{Conclusion}
In this paper, we presented a DETERlab-based IoT botnet testbed built within a secure contained environment with ancillary services such as DHCP, DNS and botnet infrastructure comprising of CnC and scanListen/loading servers. We described the setup process for various components of the testbed and associated challenges. The main features of the testbed were also listed. Finally, we ran some basic experiments and discussed the results showing bot-CnC communication, scanning traffic and few DDoS attacks which demonstrated the capabilities of our testbed.     
\section*{Acknowledgment}
The authors would like to appreciate the National Cybersecurity R\&D Lab, Singapore for allowing us to use their testbed to collect important data which has been used in our work. This research is supported by the National Research Foundation, Prime Minister’s Office, Singapore under its Corporate Laboratory@University Scheme, National University of Singapore, and Singapore Telecommunications Ltd.

\bibliographystyle{splncs}
\begingroup
\raggedright
\bibliography{oqeprop}
\endgroup

\end{document}